# A Closed-loop controller to improve the Stability of Cascaded DC/DC Converters

Maziar Isapour Chehardeh, Student member, IEEE

isapour@siu.edu

Ehsan. M. Siavashi, Student member, IEEE

e3mohamm@uwaterloo.ca

*Abstract-* **Study of the buck converter and cascaded system considering the voltage mode controller has been done. First the small signal analysis of a buck dc/dc converter is presented and its mathematical representation has been showed. Then, the cascaded converter model regarding close loop impedances and voltage gain has been studied. The controller for this converter is proposed to stabilize the performance of the plant. The effectiveness of the proposed controller has been tested on a typical buck converter.**

## I. INTRODUCTION

Power electronics has been a key factor in the rapid development of advanced vehicular systems such as land, sea/undersea, air, and space vehicles [1]. Real power system contains sub-systems which are designed to meet some specific goals without considering the existence of possible problems related to applying them in a network. On the other hand, when the sub-systems has been considering as a united network, instability may be expected [2][3]. The stability problem is one of the intriguing areas for researchers in academia and industry because of its high importance for the reliability of a dc network. [4]. Because of this issue, the study of stability in dc power supply distribution systems is important and received much attention among researchers [3][5][6][7].

In this paper, at first, the single buck converter with and without voltage mode controller (closed and opened loop) has been studied and small signal for them has been done. The average model of converter is derived and based on the derived transfer functions and impedances, stability issues for cascaded converter model has been studied. In this case, the whole system may be unstable when the impedance mismatching as the study of Middlebrook [8]-[13].

## II. SINGLE BUCK CONVERTER

The schematic of a single buck converter is illustrated as Fig.1. The average modeling of the converter can be derived. In this model, $v_{in}$ is the small-signal of the input voltage, $d$ is the small-signal of the duty cycle, $i_{load}$ is the small signal current that comes from the output and $v_c$ is the small-signal of the output voltage. The following formulas show the model of the single buck converter:

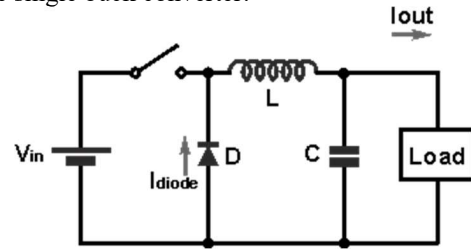

Fig.1 Single Buck converter

Regarding above equations, the dynamic model of the buck converter can be derived as bellow

$$\frac{di_l}{dt} = \frac{1}{L} \times (d \times v_{in} - v_c) \qquad (1)$$

$$\frac{dv_c}{dt} = \frac{1}{C} \times \left(i_l - \frac{1}{R}v_c - i_{load}\right) \qquad (2)$$

$$i_{in} = d \times i_l \qquad (3)$$

where, $i_l$ and $v_c$ are considered as state variables. Respectively, input voltage, load current, and duty cycle are the inputs and $i_l$, $v_c$, and input current are applied as outputs.

$$\begin{aligned} x^o &= Ax + Bu \\ y &= Cx + Du \end{aligned} \qquad (4)$$

$$x = \begin{bmatrix} i_l \\ v_c \end{bmatrix} \quad y = \begin{bmatrix} i_l \\ v_c \\ i_{in} \end{bmatrix} \quad u = \begin{bmatrix} v_{in} \\ i_{load} \\ d \end{bmatrix} \quad (5)$$

$$A = \begin{bmatrix} 0 & -\frac{1}{L} \\ \frac{1}{C} & -\frac{1}{RC} \end{bmatrix} \quad B = \begin{bmatrix} \frac{D}{L} & 0 & \frac{V_{in}}{L} \\ 0 & -\frac{1}{C} & 0 \end{bmatrix} \quad (6)$$

$$C = \begin{bmatrix} 1 & 0 \\ 0 & 1 \\ d & 0 \end{bmatrix}$$

$$H(s) = C(SI - A)^{-1}B \quad (7)$$

Considering Fig.2 and (7), the open-loop transfer functions can be derived as follow

$$H(s) = \begin{bmatrix} \frac{D(RCS+1)}{LRCS^2+LS+R} & \frac{R}{LRCS^2+LS+R} & \frac{V_{in}(RCS+1)}{LRCS^2+LS+R} \\ \frac{RD}{LRCS^2+LS+R} & \frac{-LRS}{LRCS^2+LS+R} & \frac{RV_{in}}{LRCS^2+LS+R} \\ \frac{D^2(RCS+1)}{LRCS^2+LS+R} & \frac{RD}{LRCS^2+LS+R} & \frac{DV_{in}(RCS+1)}{LRCS^2+LS+R} \end{bmatrix} \quad (8)$$

$$G_{vgol} = \frac{v_c(s)}{v_{in}(s)} = H_{21}(s) \quad (9)$$

$$= Z_{inol} = \frac{v_{in}(s)}{i_{in}(s)} = \frac{1}{H_{31}(s)} \quad (10)$$

$$Z_{outol} = -\frac{v_c(s)}{i_{load(s)}} = -H_{22}(s) \quad (11)$$

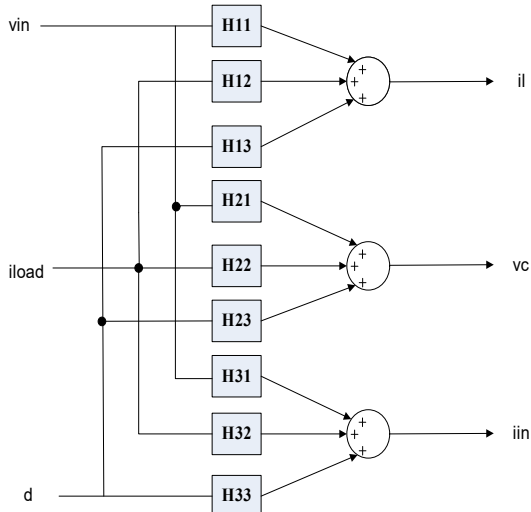

Fig.2 Block diagram of the open-loop buck converter

Here a voltage controller is connected to a system which is illustrated in Fig. 3. In this situation the output voltage $v_c$ is compared with a reference value and errors goes to a PI controller and makes the duty cycle.

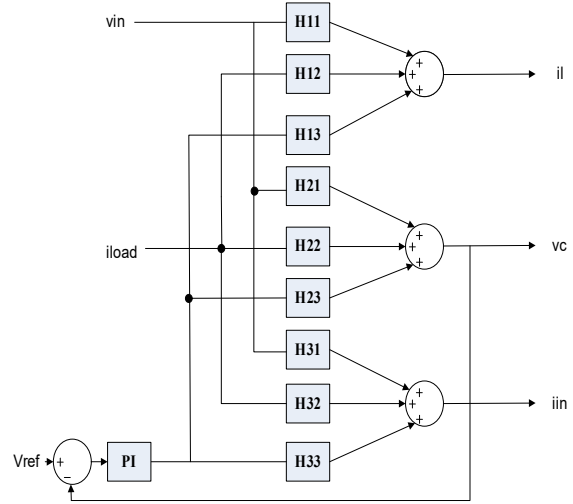

Fig. 3 Block Diagram of the Closed-Loop Buck Converter

The transfer function for above closed-loop system can be written as bellow:

$$i_L(s) = H_{11}v_{in(s)} + H_{12}i_{load(s)} + H_{13}\,d(s)$$
$$v_C(s) = H_{21}v_{in(s)} + H_{22}i_{load(s)} + H_{23}\,d(s)$$
$$i_{in}(s) = H_{31}v_{in(s)} + H_{32}i_{load(s)} + H_{33}\,d(s) \quad (12)$$
$$d(s) = G_{c(s)}(v_{ref(s)} - v_C(s))$$

$$G_{vgcl} = \frac{v_C(s)}{v_{in(s)}} = \frac{H_{21}}{[1 + H_{23}\,G_{c(s)}]} \quad (13)$$

$$Z_{incl}(s)^{-1} = \frac{i_{in}(s)}{v_{in(s)}}$$
$$= H_{31} - H_{33}\,G_{c(s)}\frac{H_{21}}{[1+H_{23}\,G_{c(s)}]} \quad (14)$$

$$Z_{outcl}(s) = -\frac{v_C(s)}{i_{load(s)}} = -\frac{H_{22}}{[1 + H_{23}\,G_{c(s)}]} \quad (15)$$

With respect to above transfer function, bode plot of the converter for two scenarios (with and without controller) can be derived. In order to have the stable system and removing the harmonic, it is useful to have large input impedance and small output impedance. The bode plots show the controller help to meet these situations.

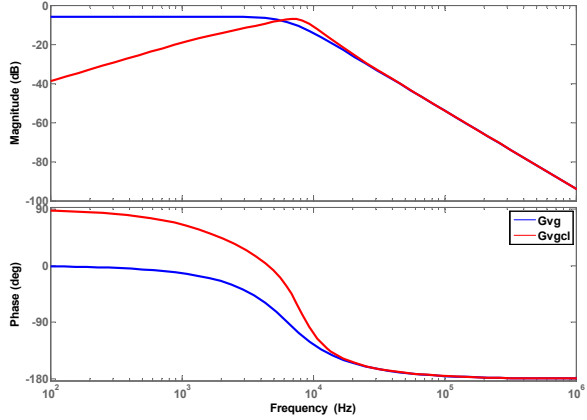

Fig. 4  Bode diagram of voltage gain

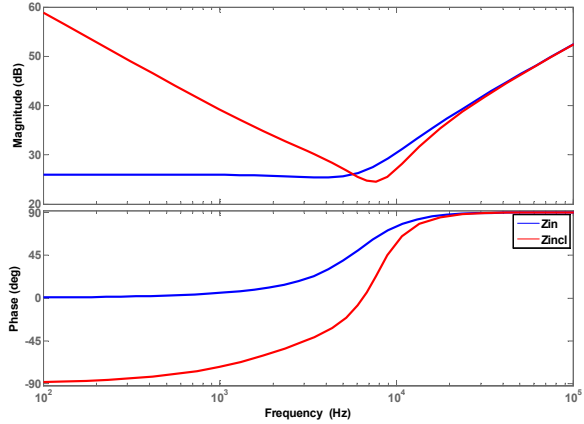

Fig. 5  Bode diagram of input impedance

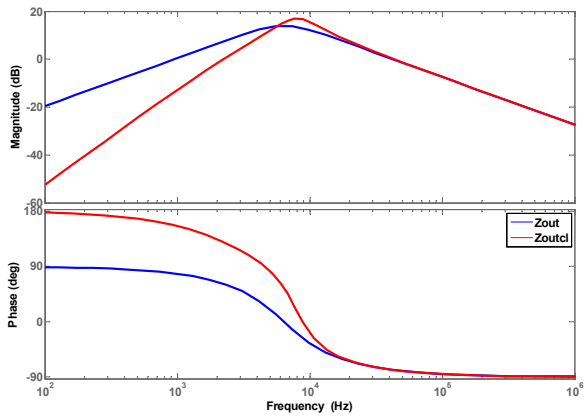

Fig. 6.  Bode diagram of output impedance

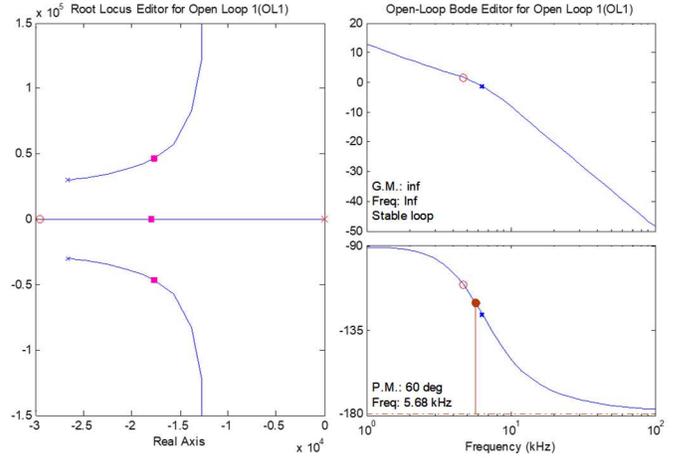

Fig. 7  Designing of controller for the converter

The matlab tool box is used to design the PI controller. The PI controller has following parameters:

Kp= 0.0093602
Ki=275.3

### III. CASCADE CONVERTER

From section II, it can be derived that single buck converter is stable and controller helps the system to be more stable. In this section two buck converters are connected together as Fig. 8.

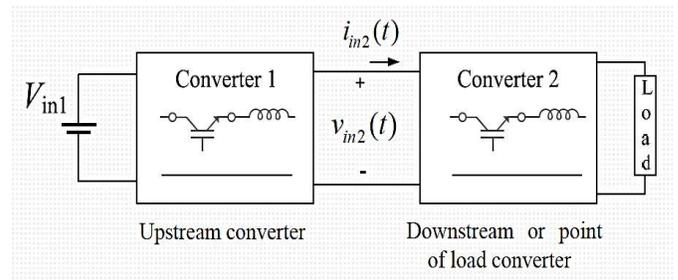

Fig. 8 Cascaded converter systems

The transfer functions of the cascaded model can be derived regarding the block diagram Fig. 9 as follow

$$v_{c1}(s) = v_{in2(s)} = H_{21}v_{in1(s)} + H_{22}i_{lin2(s)} \quad (16)$$
$$+ H_{23}\, G_{c(s)}\left(v_{ref(s)} - v_{in2}(s)\right)$$

$$v_{in2}(1 + H_{23}G_c) = H_{21}v_{in1(s)} + H_{22}i_{lin2(s)} \quad (17)$$

$$v_{C2}(s) = H'_{21}v_{in2(s)} + H'_{23}\, G'_{c(s)}(-v_{c2}(s)) \quad (18)$$

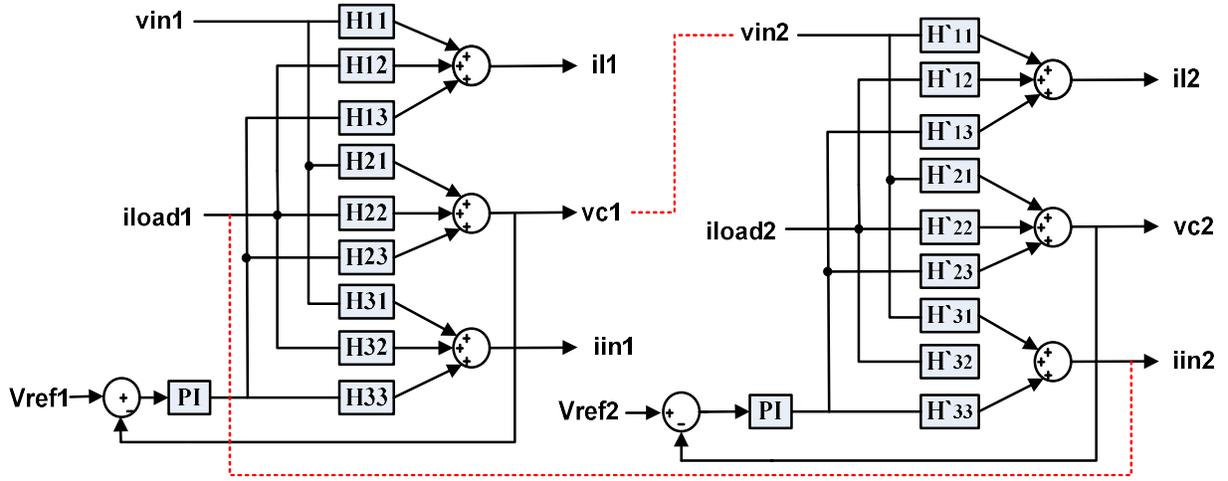

Fig. 9. closed-loop cascaded converter

$$v_{C2}(1 + H'_{23} G'_{c(s)}) = H'_{21} v_{in2(s)} \quad (19)$$

$$i_{in2}(s) = \frac{H'_{31}}{G_{vgcl2}} v_{c2} - H'_{33} G'_c v_{c2} \quad (20)$$

$$i_{in2}(s) = \left[\frac{H'_{31} v_{in2(s)} - H'_{33} G'_{c(s)} G_{vg}}{G_{vg}}\right] v_{C2} \quad (21)$$

$$i_{in2}(s) = \left[\frac{1}{G_{vgc}} Z_{incl}(s)^{-1}\right] v_{C2} \quad (22)$$

$$v_{in2}(1 + H_{23} G_c) = H_{21} v_{in1(s)} + H_{22} \left[\frac{1}{G_{vg}} Z_{incl}(s)^{-1}\right] v_{C2} \quad (23)$$

$$G_{vgt} = \frac{v_{C2}}{v_{in1}} = \frac{G_{vgcl2} \times G_{vgcl1}}{1 + \frac{Z_{outcl1}}{Z_{incl2}}} \quad (24)$$

Regarding the Middlebrook study, for having stable system following criterion should be considered:

$$\left|\frac{Z_{outcl1}}{Z_{incl2}}\right| < 1 \quad (25)$$

## IV. CASE STUDY

In this section, the buck converter and cascade system are simulated by MATLAB/PLECS software. Table I shows the parameter which are used in simulation.

TABLE I
Parameter of case study

| $V_{in1}=V_{in2}$ | 100 v | $L_1$ | 0.167 mH |
|---|---|---|---|
| $C_1$ | 3.75 uF | $R_1$ | 5 |
| $L_2$ | 0.003 mH | $C_2$ | 23.44 uF |
| $R_2$ | 0.8 | | |

In first scenario, each converter has been simulated separately to show the operation of the buck converter. Then the cascaded converter has been studied. It is worth noting that in second scenario $R_1$ is ignored.

Fig. 10, 11, and 12 show the output voltage of the first, second and cascaded converters respectively.

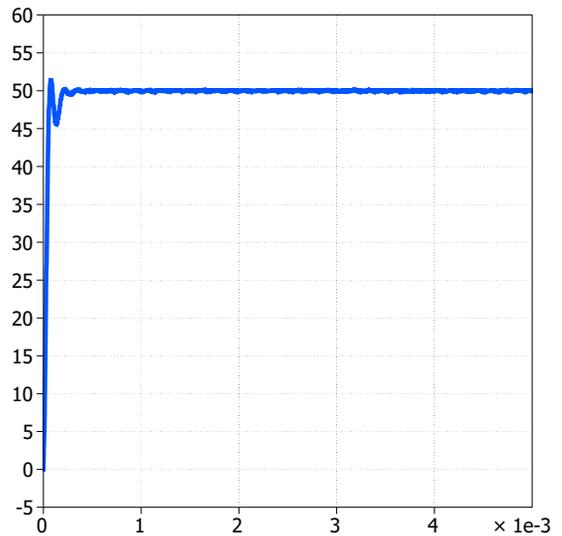

Fig. 10. output voltage of first buck converter

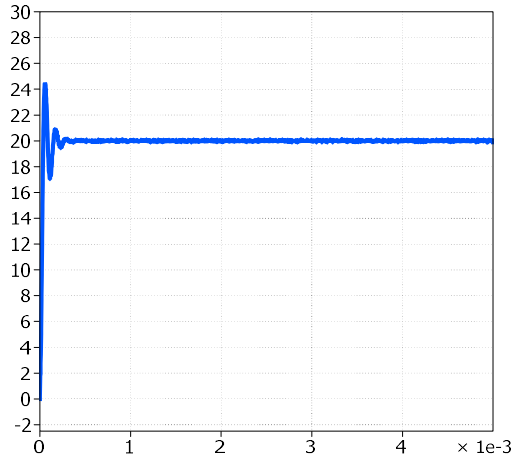

Fig. 11. output voltage of second buck converter

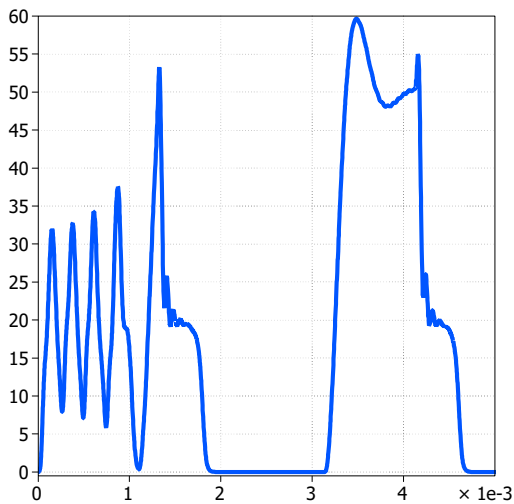

Fig. 12. output voltage of cascaded converter

Based on the mentioned systems, parameter of voltage controller are as bellow

TABLE II
The parameters of the controller

| $K_{p1}$ | 0.0093602 | $K_{i1}$ | 275.3 |
|---|---|---|---|
| $K_{p2}$ | 0.01956 | $K_{i2}$ | 537.4 |

Fig. 10 and 11 shows that each converter is stable if they work separately, however the cascaded system is not stable.

V. Conclusion

The aim of this report was study of stability of buck converters. When the converters work separately, system is stable and controller helps them to work more reliable. But if two buck converters form cascaded converter we cannot expect to have stable system, even they are stable separately.

REFERENCES

[1] S. D. Sudhoff, S. F. Glover, P. T. Lamm, D. H. Schmucker, and D. E. Delisle, "Admittance space stability analysis of power electronic systems," Aerospace and Electronic Systems, IEEE Transactions on, vol. 36, pp. 965-973, 2000.

[2] Chehardeh, M. Isapour, et al. "An optimal control strategy to alleviate sub-synchronous resonance in VSC-HVDC systems." *Power Electronics and Intelligent Transportation System (PEITS), 2009 2nd International Conference on*. Vol. 1. IEEE, 2009.

[3] R. Ahmadi, D. Paschedag, and M. Ferdowsi, "Analyzing Stability Issues in a Cascaded Converter System Comprised of Two Voltage-Mode Controlled DC-DC Converters," in Proc. IEEE Appl. Power Electron. Conf., Mar. 2011, pp. 1769-1775

[4] Hyoung Y. Cho, E. Santi, "Modeling and Stability Analysis of Cascaded Multi-Converter Systems including Feedforward and Feedback Control" Industry Application Society, 2008.

[5] R. D. Middlebrook, "Input filter considerations in design and application of switching regulators," Conf. Rec. IEEE IAS Annu. Meeting, pp. pp. 366-382, 1976.

[6] R. Ahmadi, D. Paschedag, and M. Ferdowsi, "Closed-loop input and output impedances of dc-dc switching converters operating in voltage and current mode control," in Proc. IEEE Ind. Electron. Conf., Nov. 2010, pp. 2311-2316

[7] R. Ahmadi, and M. Ferdowsi, "Improving Performance of a DC-DC Cascaded Converter System Using an Extra Feedback Loop," accepted for publication ECCE 2013.

[8] Almalki, Mishari Metab, Maziar Isapour Chehardeh, and Constantine J. Hatziadoniu. "Capacitor bank switching transient analysis using frequency dependent network equivalents." *North American Power Symposium (NAPS), 2015*. IEEE, 2015.

[9] S. Sajadian and E. C. dos Santos, "Three-phase DC-AC converter with five-level four-switch characteristic," *2014 Power and Energy Conference at Illinois (PECI)*, Champaign, IL, 2014, pp. 1-6.

[10] M. I. Chehardeh, M. M. Almalki and C. J. Hatziadoniu, "Remote feeder transfer between out-of-phase sources using STS," *2016 IEEE Power and Energy Conference at Illinois (PECI)*, Urbana, IL, 2016, pp. 1-5.

[11] Q. Xie, A. Yekkehkhany and Y. Lu, "Scheduling with multi-level data locality: Throughput and heavy-traffic optimality," *IEEE INFOCOM 2016 - The 35th Annual IEEE International Conference on Computer Communications*, San Francisco, CA, 2016, pp. 1-9.

[12] Mortezapour, Vahid, et al. "Adaptive control of chaotic ferroresonant oscillations in electromagnetic voltage transformers." *24th International Power System Conference*. 2009.

[13] A. Parizad, A. H. Khazali and M. Kalantar, "Sitting and sizing of distributed generation through Harmony Search Algorithm for improve voltage profile and reducuction of THD and losses," *CCECE 2010*, Calgary, AB, 2010, pp. 1-7